\documentclass{PoS}
\usepackage[pdftex]{graphicx}
\usepackage[cp1251]{inputenc}
\usepackage[T2A]{fontenc}
\usepackage[english,russian]{babel}
\usepackage{amsmath}
\usepackage[russian]{babel}
\usepackage{amsmath}
\usepackage{amssymb,amsfonts,amsmath,amsbsy,mathtext,cite,enumerate,float, bm}
\usepackage{graphicx}
\title{Neutrino electromagnetic properties: new approach to oscillations in magnetic fields}

\ShortTitle{New approach to neutrino oscillations in magnetic fields}

\author{Alexander Dmitriev\\
        Affiliation\\
        Department of Theoretical Physics, Faculty of Physics, Lomonosov Moscow State University, Moscow
        119991, Russia\\
        E-mail: \email{ai.dmitriev@physics.msu.ru}}

\author{Riccardo Fabbricatore\\
        Affiliation\\
        Department of Theoretical Physics, Faculty of Physics, Lomonosov Moscow State University, Moscow
        119991, Russia\\

        E-mail: \email{riccardo.fabbricatore1@gmail.com}}

        \author{\speaker{Alexander Studenikin}\\
        Affiliation\\
        Department of Theoretical Physics, Faculty of Physics, Lomonosov Moscow State University, Moscow 119991, Russia\\
        Joint Institute for Nuclear Research, Dubna 141980, Moscow Region, Russia\\
        E-mail: \email{studenik@srd.sinp.msu.ru}}

\abstract{Several new and interesting aspects of neutrino oscillations in a magnetic field are considered:
$1)$ we develop a standard usually used approach to the neutrino spin oscillations in the neutrino mass basis and obtain the effective neutrino spin (and ``spin-mass'') oscillation Hamiltonian that can be used for description of the neutrino oscillations between different pairs of neutrino states with different masses and helicities;  $2)$ we derive the exact solution of the Dirac equation for a massive neutrino with nonzero magnetic moment in the presence of a constant transversal magnetic field that is rotating along the direction of the neutrino propagation (the twisting magnetic field) and on the basis of the obtained energy spectrum the neutrino spin oscillation effective Hamiltonian is derive; $3)$ we develop a new approach to neutrino spin oscillations that is based on the description of the neutrino spin states with the corresponding spin operator that commutes with the neutrino dynamics Hamiltonian in the magnetic field. The obtained new results can have important phenomenological applications.}

\FullConference{Proceedings of the Corfu Summer Institute 2014 \\

                 3-21 September 2014\\

                 Corfu, Greece}

\begin{document}

\section{Introduction}
Experimental and theoretical studies of flavour
conversion in solar, atmospheric, reactor and accelerator neutrino
fluxes give strong evidence for nonzero neutrino masses.  A massive
neutrino can have nontrivial electromagnetic properties and, in particular, nonzero magnetic moment \cite{Fujikawa:1980yx}. A detailed review on the neutrino electromagnetic interactions is given in \cite{Giunti:2014ixa}. If neutrinos have   magnetic moments then in the presence of a magnetic field neutrinos spin can process \cite{Cisneros:1970nq,Fujikawa:1980yx,Schechter:1981hw,Voloshin:1986ty,Okun:1986hi,Dar:1987yv}.
The possibility for the resonance amplification of the neutrino spin-flavour procession (if flavour neutrinos have the so-called transition magnetic moment) was first considered in \cite{Akhmedov:1988uk,Lim:1987tk}.

Neutrino transitions and oscillations in the presence of a magnetic field have been investigated before in different aspects. A detailed review of this topic is presented in \cite{Giunti:2014ixa}  where the corresponding references to numerous publications can be also found. In our present paper we discuss several important issues in theoretical description of neutrino spin and spin-flavour oscillations.
In Section \ref{Sec_2}  we develop a standard usually used approach to the neutrino spin oscillations in the neutrino mass basis and obtain the effective neutrino spin (and ``spin-mass'') oscillation Hamiltonian that can be used for description of the neutrino oscillations between different pairs of neutrino states with different masses and helicities. In Section \ref{Sec_3} we derive the exact solution of the Dirac equation for a massive neutrino with nonzero magnetic moment in the presence of a constant transversal magnetic field that is rotating along the direction of the neutrino propagation (the twisting magnetic field). Using the obtained energy spectrum the neutrino spin oscillation effective Hamiltonian is derived. In Section \ref{Sec_4} we develop a new approach to neutrino spin oscillations that is based on the description of the neutrino spin states with the corresponding spin operator that commutes with the neutrino dynamics Hamiltonian in the magnetic field. In the customary approach the neutrino helicity operator is used that does not commute with the neutrino dynamics Hamiltonian in the presence of a magnetic field.

\section{Standard approach to neutrino spin oscillations in a magnetic field}
\label{Sec_2}

Consider two physical neutrinos, $\nu_1$ and $\nu_2$, with corresponding masses $m_1$ and $m_2$. Each of the physical neutrinos are a superposition of the neutrino flavour states, $\nu_{e}$ and $\nu_{\mu}$. In general, a neutrino beam is a superposition of the mass states $\nu_1$ and $\nu_2$, or alternatively of the flavour states
$\nu_{e}$ and $\nu_{\mu}$. Usually a notation for two neutrino basis are used,
\begin{equation}\label{nu_p_nu_f}
	\nu^{\left(p\right)}=\begin{pmatrix} \nu_{1} \\ \nu_{2} \end{pmatrix}, \ \  \ \
\nu^{\left(f\right)}=\begin{pmatrix} \nu_{e} \\ \nu_{\mu} \end{pmatrix},
\end{equation}
and two neutrino basis are related by the mixing matrix
\begin{equation}\label{mixing}
	\nu^{\left(f\right)}=U\nu^{\left(p\right)},   \ \ \
	U=\begin{pmatrix}\cos\theta & \sin\theta \\ -\sin\theta & \cos\theta \end{pmatrix},
\end{equation}
where $\theta$ is the mixing angle. We restrict ourselves to two neutrino generations. The evolution of the neutrino beam in the physical basis is governed by the Schr$\ddot{o}$dinger type equation
 \begin{equation}\label{schred_eq}
	i\frac{d}{dt} \nu^{ \left( p \right) }\left(t\right)=H_{vac}\nu^{ \left( p \right) }\left(t\right),
\end{equation}
where the Hamiltonian
\begin{equation}\label{vac_ham}
	H_{vac}=\begin{pmatrix} E_{1} & 0 \\ 0 & E_{2}  \end{pmatrix}
\end{equation}
is diagonal in this basis, and the relativistic neutrino energies are
\begin{equation}\label{energ}
	E_{\alpha} = \sqrt{\mathbf{p}^2 + m_{\alpha}^2} \approx |\mathbf{p}|+\frac{m_{\alpha}^{2}}{2|\mathbf{p}|}, \ \ \ \alpha =1,2.
\end{equation}

Consider neutrino oscillations in the presence of a constant magnetic field given by the vector $\bm {B}$. As it has been already mentioned in the introduction, a massive neutrino should have a nonzero magnetic moment. We introduce notations $\mu_{11}$ and $\mu_{22}$ for the corresponding magnetic moments of neutrinos $\nu_1$ and $\nu_2$. The magnetic moments interact with a magnetic field and from classical physics we know that the magnetic field exerts a torque on the magnetic moment. In quantum field theory, magnetic moment interaction between two neutrino fields $\nu_{\alpha}$ and $\nu_{\alpha'}$ ($\alpha '=1, 2 $) is given by
\begin{equation}\label{em_hamilt}
	H_{EM} = -\frac{1}{2}\mu_{\alpha \alpha '}\overline{\nu}_{\alpha '}\sigma_{\mu \nu}\nu_{\alpha}F^{\mu \nu} + h.c. \ ,
\end{equation}
$F^{\mu \nu}$ is the electromagnetic field tensor. In a constant magnetic field the Hamiltonian (\ref{em_hamilt}) is
\begin{equation}\label{B_hamilt}
	H_{EM} = -\frac{1}{2}\mu_{\alpha \alpha '}\overline{\nu}_{\alpha '}\bm{\Sigma B}\nu_{\alpha} + h.c. ,
\end{equation}
here
\begin{equation}
\label{Sigma}
\Sigma_i=\left(\begin{array}{cc}
                   \sigma_i & 0 \\
                   0 & \sigma_i
                 \end{array}\right),
\end{equation}
and
$\sigma_i$ are the Pauli matrices.
 When $\nu_{\alpha}$ and $\nu_{\alpha '}$ represents different fields ($\alpha \neq\alpha '$) the values $\mu_{\alpha \alpha '}$ are called the transition magnetic moments.

Each of the neutrinos $\nu_1$ and $\nu_2$ can be in two spin states. These two neutrino quantum spin states with the same mass can be considered as two different neutrino species. Totally there are four neutrino species that can be mixed by the magnetic field presence and transitions between different neutrino spin states $\nu_{\alpha,s}\Leftrightarrow\nu_{\alpha ',s'}$ ($\alpha,\alpha '= 1,2, \ s\neq s'$) can be generated. Here $s$ and $s'$ are two different neutrino spin quantum numbers. The probability of these conversions depend on time that provides the existence of neutrino spin oscillations.  Note that there are two different types of neutrino spin oscillations that precede without changing of a neutrino mass state ($\alpha=\alpha '$) and with changing mass state ($\alpha\neq\alpha '$). The later case we call the neutrino spin-mass oscillations.

When a neutrino with nonzero magnetic moment is placed in an external magnetic field, different spin states acquire different magnetic potential energies which are similar to the matter potential \cite{Wolfenstein:1977ue} and must be accounted in the effective Hamiltonian in the Schr$\ddot{o}$dinger type neutrino evolution equation (\ref{schred_eq}) (see also \cite{Giunti:2014ixa}).

\begin{center}
%\begin{minipage}[h]{1\linewidth}
\begin{figure}[h]
\center{\includegraphics[width=0.3\linewidth]{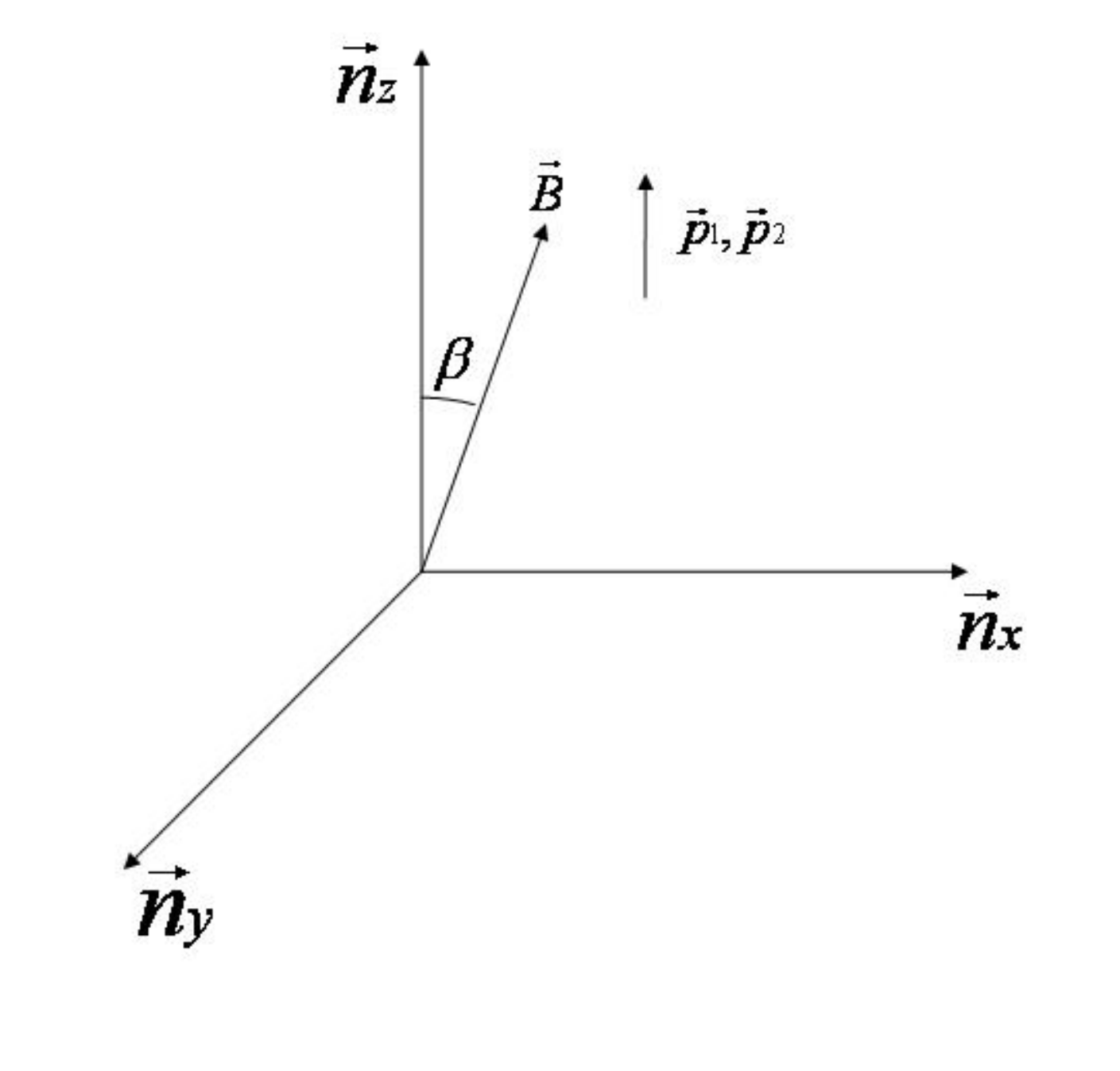}}
\vspace{-0.5cm}
\caption{An external magnetic field  ${\bf B}={\bf B}_{\perp}+{\bf B}_{\parallel}$ orientation
in respect to the neutrino $\nu_{1}$ and $\nu_{2}$ momenta ${\bf p}_1$ and ${\bf p}_2$.}
\label{figure_B}
\end{figure}
%\end{minipage}
\end{center}

Let us calculate the  magnetic potential energies $V_{\alpha,s; \alpha ', s'}$ corresponding to transitions between different neutrino states in the presence of an arbitrary oriented in respect to the neutrino propagation magnetic field. It is supposed (see Fig. \ref{figure_B}) that neutrinos momenta are along  $\bf{n}_{z}$ axis and ${\bf B}_\perp$ and ${\bf B}_{\parallel}$ are the transverse and longitudinal components of the magnetic field.

From the magnetic field interaction Hamiltonian (\ref{B_hamilt}) it follows that
\begin{equation}\label{V}
V_{\alpha,s; \alpha ', s'}=\langle	{\nu_{\alpha,s}}|H_{EM}|{\nu_{\alpha ', s'}}\rangle=-\frac{\mu_{\alpha, \alpha '}}{2}\int{d^4x \nu_{\alpha}^{\dagger}\gamma_0\begin{pmatrix}\bm{\Sigma B} & 0 \\ 0 & \bm{\Sigma B}\end{pmatrix}\nu_{\alpha '}},
\end{equation}
where
\begin{equation}	 \nu_{\alpha,s}=C_{\alpha}\sqrt{\frac{E_{\alpha}+
m_{\alpha}}{2E_{\alpha}}}\begin{pmatrix}u_{s} \\ \frac{\bm{\Sigma p_{\alpha}}}{E_{\alpha}+m_{\alpha}}u_{s}\end{pmatrix}e^{-iE_{\alpha}t+i\bm{p_{\alpha} x}},
\end{equation}
and
\begin{equation}
	\nu_{\alpha ',s'}\left(m_{\alpha '},p_{\alpha '},E_{\alpha '}\right)=C_{\alpha '}\sqrt{\frac{E_{\alpha '}+m_{\alpha '}}{2E_{\alpha '}}}\begin{pmatrix}u_{s'} \\ \frac{\bm{\Sigma p_{\alpha '}}}{E_{\alpha '}+m_{\alpha '}}u_{s'}\end{pmatrix}e^{-iE_{\alpha '}t+i\bm{p_{\alpha '} x}}.
\end{equation}

For the considered case of ultrarelativistic neutrinos we describe neutrino spin states by their helicities. Therefore for a particular type of a massive neutrino $\nu_{\alpha}$ there are two different spin states given by  \begin{equation}\label{u_s_plus}
	u_{s=1}=\begin{pmatrix}1 \\ 0\end{pmatrix}, \ \ \ \
	u_{s=-1}=\begin{pmatrix}0 \\ 1\end{pmatrix},
\end{equation}
that correspond to the right-handed $\nu_{R}$ and left-handed $\nu_{L}$ chiral neutrinos, respectively. As well known \cite{Cisneros:1970nq,Fujikawa:1980yx,Voloshin:1986ty,Okun:1986hi}, the spin of such a neutrino can precess in a transversal magnetic field.

From (\ref{V}) we get

\begin{multline}V_{\alpha,s; \alpha ', s'}=-\frac{1}{2}\mu_{{\alpha \alpha '}}
C_{\alpha}C_{\alpha '}\int{d^4x\bm{B}\begin{pmatrix}u_{s}^{\dagger} & \frac{\bm{\Sigma p_{\alpha}}}{E_{\alpha}+m_{\alpha}}u_{s}^{\dagger}\end{pmatrix} \begin{pmatrix}\bm{\Sigma} & 0 \\ 0 & -\bm{\Sigma}\end{pmatrix}\begin{pmatrix}u_{s'} \\ \frac{\bm{\Sigma p_{\alpha '}}}{E_{\alpha '}+m_{\alpha '}}u_{s'}\end{pmatrix}}\\ \\ \times \frac{\sqrt{\left(E_{\alpha}+m_{\alpha}\right)\left(E_{\alpha '}+m_{\alpha '}\right)}}{2\sqrt{E_{\alpha}E_{\alpha '}}} exp\left(-i\Delta Et+i\Delta px\right) .\end{multline}
It is possible to show that
\begin{multline}\bm{B}\begin{pmatrix}u_{s}^{\dagger} & \frac{\bm{\Sigma p_{\alpha}}}{E_{\alpha}+m_{\alpha}}u_{s}^{\dagger}\end{pmatrix} \begin{pmatrix}\bm{\Sigma} & 0 \\ 0 & -\bm{\Sigma}\end{pmatrix}\begin{pmatrix}u_{s'} \\ \frac{\bm{\Sigma p_{\alpha '}}}{E_{\alpha '}+m_{\alpha '}}u_{s'}\end{pmatrix}=\\ \\=u_{s}^{\dagger} \left( \left(\bm{\Sigma B_{||}}\right)\left(1-\frac{\bm{p_{\alpha} p_{\alpha '}}}{\left(E_{\alpha}+m_{\alpha}\right)\left(E_{\alpha '}+m_{\alpha '}\right)}\right)+\bm{\Sigma B_{\perp}}\left(1+\frac{\bm{p_{\alpha} p_{\alpha '}}}{\left(E_{\alpha}+m_{\alpha}\right)\left(E_{\alpha '}+m_{\alpha '}\right)} \right) \right)  u_{s'}.\end{multline}
In the case of ultrarelativistic neutrinos $\frac{m_{\alpha}}{E_{\alpha}} \ll 1$ we have
\begin{equation}\left(1-\frac{\bm{p_{\alpha} p_{\alpha '}}}{\left(E_{\alpha}+m_{\alpha}\right)\left(E_{\alpha '}+m_{\alpha '}\right)}\right)\frac{\sqrt{\left(E_{\alpha}+m_{\alpha}\right)\left(E_{\alpha '}+m_{\alpha '}\right)}}{2\sqrt{E_{\alpha}E_{\alpha '}}}=\gamma^{-1}_{\alpha \alpha '},\end{equation}
where
\begin{equation}
	\gamma^{-1}_{\alpha \alpha '} = \frac{1}{2}\left(\frac{m_{\alpha}}{E_{\alpha}}+\frac{m_{\alpha '}}{E_{\alpha '}}\right).
\end{equation}
and
\begin{equation}\left(1+\frac{\bm{p_{\alpha} p_{\alpha '}}}{\left(E_{\alpha}+m_{\alpha}\right)\left(E_{\alpha '}+m_{\alpha '}\right)}\right)\frac{\sqrt{\left(E_{\alpha}+m_{\alpha}\right)\left(E_{\alpha '}+m_{\alpha '}\right)}}{2\sqrt{E_{\alpha}E_{\alpha '}}}=1.\end{equation}
In the strightforward calculations we obtain that
\begin{multline}u^{\dagger}_{s=1}\bm{\Sigma B}u_{s=1}=\begin{pmatrix}1 & 0\end{pmatrix}\bm{\Sigma}\begin{pmatrix}1 \\ 0\end{pmatrix}\left(\bm{B_{||}}+\bm{B_{\perp}}\right)=\\=\begin{pmatrix}1 & 0\end{pmatrix}\sigma_3\begin{pmatrix}1 \\ 0\end{pmatrix}B\cos \beta + \begin{pmatrix}1 & 0\end{pmatrix}\sigma_1\begin{pmatrix}1 \\ 0\end{pmatrix}B\sin \beta=B\cos \beta ,\end{multline}
\begin{equation}u^{\dagger}_{s=1}\bm{\Sigma B}u_{s=-1}=B\sin \beta ,\end{equation}
\begin{equation}u^{\dagger}_{s=-1}\bm{\Sigma B}u_{s=1}=B\sin \beta ,\end{equation}
\begin{equation}u^{\dagger}_{s=-1}\bm{\Sigma B}u_{s=-1}=-B\cos \beta .\end{equation}
As it can be expected, in neutrino transitions without change of helicity only the $B_{\parallel}=B\cos \beta$ component of the magnetic field contribute to the effective potential, whereas in transitions with change of the neutrino helicity the transversal component $B_{\perp}=B\sin \beta$ matters.

Now we consider two different neutrino mass states with two helicities as four neutrino species
$\nu_{1,s=1}, \ \nu_{1,s=-1}, \ \nu_{2,s=1}$ and $\nu_{2,s=-1}$. From the performed calculations we
conclude that the effective neutrino oscillation Hamiltonian that governs the massive neutrino
oscillations in the presence of a constant magnetic field is
\begin{equation}\label{schred_eq}
	H_{em}=H_{vac}+H_{\mu},
\end{equation}
where the corresponding contributions of the magnetic potentials $V_{\alpha,s; \alpha ', s'}$ to the oscillation Hamiltonian are given by
\begin{equation}H_{\mu}=
	\frac{1}{2}
	\begin{pmatrix}
	\mu_{1 1}\frac{B_{||}}{\gamma_{1 1}} & \mu_{1 1}B_{\perp} & \mu_{1 2 }
\frac{B_{||}}{\gamma_{1 2}} & \mu_{1 2 }B_{\perp} \\
	\mu_{1 1}B_{\perp} & -\mu_{1 1}\frac{B_{||}}{\gamma_{1 1}} & \mu_{1 2}B_{\perp} & -
\mu_{1 2}\frac{B_{||}}{\gamma_{1 2 }} \\
	\mu_{1 2 }\frac{B_{||}}{\gamma_{1 2 }} & \mu_{1 2 }B_{\perp} & \mu_{2  2 }\frac{B_{||}}
{\gamma_{2  2 }} & \mu_{2  2 }B_{\perp} \\
	\mu_{1 2}B_{\perp} & -\mu_{1 2}\frac{B_{||}}{\gamma_{1 2 }} & \mu_{2  2}B_{\perp} & -
\mu_{2  2 }\frac{B_{||}}{\gamma_{2  2 }}\\
	\end{pmatrix}.
	\end{equation}
Finally, for the neutrino evolution equation we get
\begin{equation}\label{gen_evol_eq}
	i\frac{d}{dt} \begin{pmatrix}\nu_{1, s=1} \\ \nu_{1, s=-1} \\  \nu_{2, s=1 } \\ \nu_{2, s=-1 }\end{pmatrix}=\frac{1}{2}
	\begin{pmatrix}
	E_{1}+\mu_{1 1}\frac{B_{||}}{\gamma_{1 1}} & \mu_{1 1}B_{\perp} & \mu_{1 2 }
\frac{B_{||}}{\gamma_{1 2}} & \mu_{1 2 }B_{\perp} \\
	\mu_{1 1}B_{\perp} & E_{1}-\mu_{1 1}\frac{B_{||}}{\gamma_{1 1}} & \mu_{1 2}B_{\perp} & -
\mu_{1 2}\frac{B_{||}}{\gamma_{1 2 }} \\
	\mu_{1 2 }\frac{B_{||}}{\gamma_{1 2 }} & \mu_{1 2 }B_{\perp} & E_{2}+\mu_{2  2 }\frac{B_{||}}
{\gamma_{2  2 }} & \mu_{2  2 }B_{\perp} \\
	\mu_{1 2}B_{\perp} & -\mu_{1 2}\frac{B_{||}}{\gamma_{1 2 }} & \mu_{2  2 '}B_{\perp} & E_{2}-
\mu_{2  2 }\frac{B_{||}}{\gamma_{2  2 }}\\
	\end{pmatrix}
	\begin{pmatrix}\nu_{1, s=1} \\ \nu_{1, s=-1} \\  \nu_{2, s=1 } \\ \nu_{2, s=-1}\end{pmatrix}.
\end{equation}

The obtained neutrino evolution equation contains information on all possible  oscillations in the presence of a magnetic field between different pairs from four neutrino states with masses $m_1$ and $m_2$ and helicities given by $s=1$ and $s=-1$. Consider three typical cases.
For neutrino evolution between two neutrino states with equal masses and different helicities, for instance  $\nu_{1,s=1}\Leftrightarrow\nu_{1,s=-1}$, from (\ref{gen_evol_eq}) it follows
\begin{equation}\label{1_evol_eq}
	i\frac{d}{dt} \begin{pmatrix}\nu_{1, s=1} \\ \nu_{1, s=-1} \\  \end{pmatrix}=\frac{\mu_{1 1}}{2}
	\begin{pmatrix}
	\frac{B_{||}}{\gamma_{1 1}} & B_{\perp}  \\
	B_{\perp} & -\frac{B_{||}}{\gamma_{1 1}}  \\
		\end{pmatrix}
	\begin{pmatrix}\nu_{1, s=1} \\ \nu_{1, s=-1} \\ \end{pmatrix}.
\end{equation}
For neutrino evolution between two neutrino states with different masses and different helicities, for instance  $\nu_{1,s=1}\Leftrightarrow\nu_{2,s=-1}$, from (\ref{gen_evol_eq}) it follows
\begin{equation}\label{2_evol_eq}
	i\frac{d}{dt} \begin{pmatrix}\nu_{1, s=1} \\ \nu_{2, s=-1} \\  \end{pmatrix}=\frac{1}{2}
	\begin{pmatrix}
	\frac{\Delta m ^2}{4E}+\mu_{1 1}\frac{B_{||}}{\gamma_{1 1}} & \mu_{1 2}B_{\perp}  \\
	\mu_{1 2}B_{\perp} & -\frac{\Delta m ^2}{4E}-\mu_{22}\frac{B_{||}}{\gamma_{2 2}}  \\
		\end{pmatrix}
	\begin{pmatrix}\nu_{1, s=1} \\ \nu_{2, s=-1} \\ \end{pmatrix}.
\end{equation}
For neutrino evolution between two neutrino states with different masses and equal helicities, for instance  $\nu_{1,s=1}\Leftrightarrow\nu_{2,s=1}$, from (\ref{gen_evol_eq}) it follows
\begin{equation}\label{3_evol_eq}
	i\frac{d}{dt} \begin{pmatrix}\nu_{1, s=1} \\ \nu_{2, s=1} \\  \end{pmatrix}=\frac{1}{2}
	\begin{pmatrix}
	\frac{\Delta m ^2}{4E}+\mu_{1 1}\frac{B_{||}}{\gamma_{1 1}} & \mu_{12}\frac{B_{||}}{\gamma_{1 2}}  \\
	\mu_{12}\frac{B_{||}}{\gamma_{1 2}} & -\frac{\Delta m ^2}{4E}+\mu_{22}\frac{B_{||}}{\gamma_{2 2}}  \\
		\end{pmatrix}
	\begin{pmatrix}\nu_{1, s=1} \\ \nu_{2, s=1} \\ \end{pmatrix}.
\end{equation}

From the obtained equations (\ref{1_evol_eq}), (\ref{2_evol_eq}) and  (\ref{3_evol_eq}) several general conclusions on the influence of an arbitrary magnetic field ${\bf B}={\bf B}_{\perp}+{\bf B}_{\parallel}$ on neutrino evolution are strightforward:

1) the mixing between two neutrinos with different helicities is due to the corresponding magnetic moment (or transition magnetic moment) interaction with the transversal magnetic field $B_{\perp}$,

2) the longitudinal magnetic field component  $B_{||}$  coupled to the corresponding magnetic moment shifts the neutrino energy,

3) in case of nonzero transition magnetic moment $\mu_{12}$, the mixing between neutrino states with different masses is induced by the interaction with $B_{||}$.

The obtained results can be translated to the flavour neutrino evolution and oscillations by inclusion of the neutrino mixing effects following to (\ref{mixing}). Then the effect of the background matter can be also added.
Note that a possibility of the resonant amplification of neutrino oscillations in the longitudinal magnetic field were discussed in \cite{Akhmedov:1988hd,Akhmedov:1990ng,Studenikin:2004bu}.

\section{Neutrino propagation and spin oscillations in twisting magnetic field}
\label{Sec_3}

The problem of neutrino spin oscillations in the presence of a transversal in respect to the neutrino propagation and twisting magnetic field, given by
\begin{equation}
{\bf{B}}_{\bot}(z)={B}_{\bot}\cos(kz){\bf{e}}_x +{B}_{\bot}\sin(kz){\bf{e}}_y,
\end{equation}
where $k=2\pi /\lambda$ and $\lambda$ is the length of a period of the magnetic field rotation,
and its possible phenomenological implications was discussed before
(see, for instance, \cite{Vidal,Smirnov,Akhmedov:1993sh,Likhachev:1990ki,Likhachev:1995xw,Studenikin:2004bu}). In this section we derive the exact solution for the Dirac equation for a massive neutrino with nonzero magnetic moment including the exact expression for the neutrino energy. Then on this basis we consider the neutrino spin and spin-mass oscillation in such a magnetic field.

We consider the Dirac equation for a massive neutrino moving in the twisting magnetic field in the following form,
\begin{equation}\label{Dir_4}
  \Big[\gamma p-m_i-\mu_{i} B_{\bot}\big(\Sigma_1\cos(kz)+\Sigma_2\cos(kz)\big)\Big]\nu_i(p)=0,
\end{equation}
where matrices are are given by (\ref{Sigma}). It is supposed that the neutrino propagates
along ${\bf e}_{z}$ direction. Then the solution can be found in the form
\begin{equation}\label{solution}
\psi(z)=Ce^{-izp_3}\begin{pmatrix}C_1e^{ikz/2}\\C_2e^{-ikz/2}\\C_3e^{ikz/2}\\
C_4e^{-ikz/2}\end{pmatrix}.\end{equation}
From (\ref{Dir_4}) and (\ref{solution}) for the neutrino wave function we obtain
the following equation
\begin{equation}
    \begin{pmatrix}
        (m_i-E)&\mu_i B_{\bot}e^{-ikz}&\hat{p}_3&0 \\
        \mu_i B_{\bot}e^{ikz}&(m_i-E)&0&-\hat{p}_3\\
        \hat{p}_3&0&-(m_i+E)&-\mu_i B_{\bot}e^{-ikz}\\
        0&-\hat{p}_3&-\mu_i B_{\bot}e^{ikz}&-(m_i+E)
    \end{pmatrix}\psi=0,
\end{equation}
and then
\begin{equation}\label{Dir_psi}
\begin{pmatrix}
    (m_i-E)&\mu_i B_{\bot}e^{-ikz}&(p_3+k/2)&0 \\
    \mu_i B_{\bot}e^{ikz}&(m_i-E)&0&-(p_3-k/2)\\
    (p_3+k/2)&0&-(m_i+E)&-\mu_i B_{\bot}e^{-ikz}\\
    0&-(p_3-k/2)&-\mu_i B_{\bot}e^{ikz}&-(m_i+E)
\end{pmatrix}\psi=0.
\end{equation}

For the neutrino energy spectrum from (\ref{Dir_psi}) we obtain
\begin{equation}
E=\pm\sqrt{m_i^2+p^2+(k/2)^2+(\mu_i B_{\bot})^2+2s\sqrt{m_i^2(\mu_i B_{\bot})^2
+p^2((\mu_i B_{\bot})^2+(k/2)^2)}},
\end{equation}
where $s=\pm 1$ correspond to two neutrino helicity states.
In the limits $p>>m_i$ , $p>>\mu_i B_{\bot}$, $p>>k$ we can simplify it to

\begin{equation}\label{nu_energy}
E=\pm\sqrt{m_i^2+(p+s\sqrt{(\mu_i B_{\bot})^2 +(k/2)^2})^2}.
\end{equation}
Finally, for the neutrino wave function components we obtain
\begin{equation}
\psi=C
\begin{pmatrix}
 1  \\
  \chi \mu_i B_{\bot} \\
  \frac{(E-m_i)+(\mu_i B_{\bot})^2\chi}{p+\frac{k}{2}}  \\
  \frac{1+(E-m_i)\chi}{p-\frac{k}{2}} \mu_i B_{\bot}
 \end{pmatrix}e^{-i(Et-(p+\frac{k}{2})z)},
\end{equation}
 where
\begin{equation}
\chi=\frac{2(pE+m_i(k/2))/(p^2-(k/2)^2)}{(p-k/2)+\frac{m_i^2-E^2}{p-k/2}-\frac{
(\mu_i B_{\bot})^2}{p+k/2}}.
\end{equation}
In the case of relativistic neutrinos from  (\ref{nu_energy}) for the energy we get
\begin{equation}\label{nu_energy_appr}
 E\approx p+\dfrac{m_i^2+(k/2)^2+(\mu_i B_{\bot})^2}{2p} \pm \sqrt{(\mu_i
  B_{\bot})^2+(k/2)^2}.
\end{equation}

Consider two neutrino species $\nu_{s=+1}$ and $\nu_{s=-1}$ with equal masses $m=m_i$ and of different spin states $s=\pm 1$  that forms a new neutrino basis
\begin{equation}\label{new_nu_basis}
	\nu=\begin{pmatrix} \nu_{s=+1} \\ \nu_{s=-1} \end{pmatrix}.\end{equation}
Obviously, the effective neutrino evolution Hamiltonian in this basis is diagonal
and the evolution equation is
\begin{equation}\label{evol_equation}
i\frac{d }{d t}\binom{\nu_{s=+1}}{\nu_{s=-1}}= \begin{pmatrix}
 +\sqrt{(\mu B_{\bot})^2+(k/2)^2}& 0\\
 0& -\sqrt{(\mu B_{\bot})^2+(k/2)^2}
\end{pmatrix} \binom{\nu_{s=+1}}{\nu_{s=-1}},
\end{equation}
here $\mu$ is the neutrino magnetic moment.
In the case of two neutrino species $\nu_{1,s=+1}$ and $\nu_{2,s=-1}$ with different masses $m_1 \neq m_2$
and different magnetic moments $\mu_1$ and $\mu_2$ the evolution equation takes the form
\begin{equation}\label{evol_equation_diff_masses}
i\frac{d }{d t}\binom{\nu_{1,s=+1}}{\nu_{2,s=-1}}= \begin{pmatrix}
 \frac{\Delta m^2 }{4p}+a_1 & 0\\
 0& -\frac{\Delta m^2}{4p}-a_2
\end{pmatrix} \binom{\nu_{1,s=+1}}{\nu_{2,s=-1}},
\end{equation}where
\begin{equation}
  a_i=\sqrt{(\mu_iB_\perp)^2+(k/2)^2}.\end{equation}

In the production and detection processes neutrino participate by their  left-handed chiral (negative helicity) states.  Therefore, in order to study possible phenomenological consequences of neutrino oscillations in twisting magnetic fields one should represent the initial left-handed neutrino as a linear combination of the basis neutrino states $\nu_{is}, \ \nu_{js'}$ ($i,j=1,2$ and $s,s'=\pm 1$), then include the neutrino oscillation process that proceeds following the neutrino evolution equation (\ref{evol_equation}) or (\ref{evol_equation_diff_masses}), and at the final stage again extract the left-handed neutrino component that only interacts in the detector. The above scheme is implemented and discussed in details in the next section.

\section{Neutrino oscillations within precise description of spin states in magnetic field}
\label{Sec_4}
In this section we show that there can be more precise approach to neutrino oscillations in the presence of a magnetic field than one usually used in literature and that we also apply in Section \ref{Sec_2}. Within this customary approach the helicity operator is used for classification of a neutrino spin states in a magnetic field. However, the helicity operator does not commute with the neutrino Hamiltonian. This case resembles situation of the flavour neutrino oscillations in the nonadiabatic case when the neutrino mass states are not stationary.  The proposed alternative approach to neutrino spin oscillations is based on the exact solutions of the corresponding Dirac equation for a massive neutrino wave function in the presence of a magnetic field that stipulates the description of the neutrino spin states with the corresponding spin operator that commutes with the neutrino dynamic Hamiltonian in the magnetic field.

Here we again consider a simple model with two generations of flavour neutrinos
$\nu_e$ and $\nu_\mu$ that are the orthogonal superpositions of mass states $\nu_1$ and $\nu_2$
\begin{equation}
\nu_f=\sum_iU_{fi}\nu_i,\end{equation}
where $U_{fi}$ are elements the mixing matrix given by (\ref{mixing}) and $f=e,\mu$, $i=1,2$.
We start with consideration of a massive $\nu_i$ with the magnetic moment $\mu_i$ that propagates  along $\bf{n}_{z}$ direction in presence of constant homogeneous arbitrary orientated magnetic field ${\bf B}=(B_\bot,0,B_\|)$. The neutrino wave function in the momentum representation is given by a plane wave solution of the modified Dirac equation
\begin{equation}\label{eq1}
  (\gamma p-m_i-{\mu_i}{\bm{\Sigma}\bf{B}})\nu_i(p)=0.
\end{equation}
The neutrino energy spectrum can be determined from the condition
\begin{equation}\det(\gamma p -m_i-{\mu_i}{\bm{\Sigma}\bf{B}})=0,\end{equation}
which guarantees the existence of a nontrivial solution of the modified Dirac equation (\ref{eq1}).
For the neutrino energy spectrum we obtain
\begin{equation}\label{spec}
E_i^\pm=\sqrt{m_i^2+p^2+{\mu_i}^2{\bf{B}}^2\pm2{\mu_i}\sqrt{m_i^2{\bf{B}}^2+p^2B_\bot^2}},
\end{equation}
where ``$\pm$'' denotes two different eigenvalues of the Hamiltonian
\begin{equation}\label{Ham_i}
  H_i=\gamma_0(m_i+{\bm{\gamma}\bf{p}}+{\mu_i}{\bm{\Sigma}\bf{B}}),
\end{equation}
which describes dynamics of the neutrino system under consideration.

Of course such a simple model can pretend to an approximate description of a
realistic physical process. It also enables one to reveal  general properties of spin and spin-flavor
oscillations in the presence of an arbitrary magnetic field that should be important for phenomenological applications.

Our approach is based, at first,  on the obtained
the exact neutrino energy spectrum (\ref{spec}) in the magnetic field.
Secondly, we define different neutrino spin states in the mass basis as eigenstates
of the spin operator (see \cite{Sokolov})
\begin{equation}\label{spin}
S_i=\frac{1}{|{\bf{B}}|}({\bm{\Sigma}\bf{B}}-\frac{i}{m_i}
\gamma_0\gamma_5\left[{\bm{\Sigma}}\times{\bf{p}}\right]{\bf{B}}),
\end{equation}
which commutes with the Hamiltonian (\ref{Ham_i}). Hence, we specify the neutrino spin states
as the stationary states for the Hamiltonian, contrary to the case when the helicity operator is used.

Consider the mass state $\nu_{i}$ as a superposition of neutrinos $\nu_i^+$ and $\nu_i^-$ in a
definite spin state,
\begin{equation}\label{super_posi}
\nu_i=c_i^+\nu_i^++c_i^-\nu_i^-.\end{equation}
The complex coefficients denote two different eigenstates of the spin operator $S_i$ and  $|c_i^+|^2+|c_i^-|^2=1$. Thus, the neutrino mass states evolve
following to
\begin{equation}\label{nui}
\nu_i(x)=\left[c_i^+e^{-iE_i^+t}\xi_i^++c_i^-e^{-iE_i^-t}\xi_i^-\right]e^{i\bm {p}\bm {x}},
\end{equation}
where the neutrino initial state at $t=0$ is given by
\begin{equation}
\nu_i^\pm (t=0)= \xi_i^\pm e^{i\bm {p}\bm {x}}.\end{equation}
In the following calculations the term $e^{i\bm {p}\bm {x}}$ is neglected because it is
irrelevant for the neutrino oscillation probability.

Next we assume that the initial neutrino state $\nu (t=0)$ is a pure electron state
which is defined as the superposition of the mass states,
\begin{equation}\label{nu_0}
\nu (t=0) =\left[c_1^+\xi_1^++c_1^-\xi_1^-\right]\cos\theta
+\left[c_2^+\xi_2^++c_2^-\xi_2^-\right]\sin\theta.
\end{equation}
Using \ref{nui} we see that this state depends on time as
\begin{equation}\label{nu_t}\nu(t)=\left[c_1^+e^{-iE_1^+t}\xi_1^++c_1^-e^{-iE_1^-t}\xi_1^-\right]\cos\theta
+\left[c_2^+e^{-iE_2^+t}\xi_2^++c_2^-e^{-iE_2^-t}\xi_2^-\right]\sin\theta.\end{equation}
Therefore, the probability to observe the muon neutrino state $\nu_\mu$ at time $t$ is given by
\begin{equation}P_{\nu_e\rightarrow\nu_\mu}(t)=\left| \left\langle \nu_\mu|\nu(t)\right\rangle \right|^2,
\end{equation}
where
$$\nu_\mu=-\left[c_1^+\xi_1^++c_1^-\xi_1^-\right]\sin\theta+\left[
c_2^+\xi_2^++c_2^-\xi_2^-\right]\cos\theta.$$

It is clear that
\begin{equation}\xi_i^{s\dag}\xi_i^{s^\prime}=\delta^
{ss^\prime} \text{and}
\ \ \
\xi_2^{s\dag}\xi_1^{s^\prime}=\xi_1^{s\dag}\xi_2^{s^\prime}=0,\end{equation}
where $s,s^\prime=\pm$. Using (\ref{nu_0}) and (\ref{nu_t}) we get
\begin{gather}
\left| \left\langle \nu_\mu|\nu(t)\right\rangle \right|=\left(|c_2^+|^2e^{-iE_2^+t}+|c_2^-|^2e^{-iE_2^-t}
-|c_1^+|^2e^{-iE_1^+t}-|c_1^-|^2e^{-iE_1^-t}\right)\sin\theta\cos\theta,
\end{gather}
and
\begin{multline}\label{P}
  P_{\nu_e\rightarrow\nu_\mu}(t)=\left\{-|c_2^+|^2|c_2^-|^2\sin^2\frac{E_2^+-E_2^-}{2}t+
  |c_2^+|^2|c_1^+|^2\sin^2\frac{E_2^+-E_1^+}{2}t+\right.\\+|c_2^+|^2|c_1^-|^2\sin^2\frac{E_2^+-E_1^-}{2}t
  +|c_2^-|^2|c_1^+|^2\sin^2\frac{E_2^--E_1^+}{2}t+\\\left.+|c_2^-|^2|c_1^-|^2\sin^2\frac{E_2^--E_1^-}{2}t
  -|c_1^+|^2|c_1^-|^2\sin^2\frac{E_1^+-E_1^-}{2}t\right\}\sin^22\theta.
\end{multline}

It is usually assumed that the initial state of relativistic neutrino is a negative-helicity state,
which means that
\begin{equation}\frac{\bm{\Sigma}\bf{p}}{|\bf{p}|}\xi_i=-\xi_i.
            \end{equation}
Next we consider the left-handed spinors because only the left-handed
fermions participate in the production and detection processes and we suppose that each of
the mass states of the initial electron neutrino are left-handed. In our case the helicity operator
is equal to ${\bm{\Sigma}\bf{p}}/{|{\bf{p}}|}=\sigma_3$, therefore the initial neutrino
state is given by
 \begin{equation}\psi_L=\left(\begin{array}{c}
                            0 \\
                            1 \\
                            0 \\
                            0
                          \end{array}\right).\end{equation}
Let us write the initial neutrino state  $\psi_L$ as a superposition of the eigenvectors of
the spin operator $S_i$. From (\ref{spin}) we get
\begin{equation}S_i=\left(\begin{array}{cccc}
              \cos\phi & \sin\phi & 0 & -\frac{p}{m_i}\sin\phi \\
              \sin\phi & -\cos\phi & \frac{p}{m_i}\sin\phi & 0 \\
              0 & \frac{p}{m_i}\sin\phi & \cos\phi & \sin\phi \\
              -\frac{p}{m_i}\sin\phi & 0 & \sin\phi & -\cos\phi
            \end{array}\right),
\end{equation}
where $\phi$ is the angle between $\bf{B}$ and $\bf{p}$. It is obvious that
\begin{equation}S_i^2=\left(1+\frac{p^2}{m_i^2}\sin^2\phi\right)\hat{I}_{4\times4}.\end{equation}
In order to define the spin projector operators we introduce the normalized spin operator following to
\begin{equation}\label{s_tilde}
\tilde{S}_i=\sqrt{\frac{1}{1+\frac{p^2}{m_i^2}\sin^2\phi}}S_i\equiv N_iS_i, \ \ \ \tilde{S}_i^2=1.
\end{equation}
The spin projector operators are
\begin{equation}P_i^\pm=\frac{1\pm \tilde{S}_i}{2},\end{equation}
and we use them to split the initial neutrino state $\psi_L$
in two neutrino states with definite spin quantum numbers
\begin{equation}\psi_i^+=P_i^+\psi_L=\frac{1}{2}\left(\begin{array}{c}
                               N_i\sin\phi \\
                               1-N_i\cos\phi \\
                               \frac{p}{m_i}N_i\sin\phi \\
                               0
                             \end{array}\right), \ \ \
\psi_i^-=P_i^-\psi_L=\frac{1}{2}\left(\begin{array}{c}
                               -N_i\sin\phi \\
                               1+N_i\cos\phi \\
                               -\frac{p}{m_i}N_i\sin\phi \\
                               0
                             \end{array}\right).\end{equation}
Note that $\psi_i=\psi_i^++\psi_i^-=c_i^+\eta_i^++c_i^-\eta_i^-,$ where $\eta_i^\pm$ is a
basis in the spin operator  $S_i$ eigenspace. From the condition
\begin{equation}\psi_i^{\dag\pm}\psi_L^\pm=|c_i^\pm|^2\end{equation}
 we get that
\begin{equation}\label{c}
|c_i^+|^2=\frac{1-N_i\cos\phi}{2},\rule{2em}{0pt}|c_i^-|^2=\frac{1+N_i\cos\phi}{2},\end{equation}
where as in eq.(\ref{s_tilde})
\begin{equation}N_i=\sqrt{\frac{1}{1+\frac{p^2}{m_i^2}\sin^2\phi}}.\end{equation}

Now we can insert the obtained expressions (\ref{c}) for $|c_i^\pm|^2$ in eq. \eqref{P}.
In the forthcoming evaluation of the probability $P_{\nu_e\rightarrow\nu_\mu}(t)$ we consider the case when the magnetic field $\bf{B}$ is nearly a transversal one and $B_\bot\gg B_\|$, therefore
\begin{equation}\sin\phi\approx1, \ \ \ \cos\phi\approx 0.\end{equation}
Then we get (it is also supposed that $\frac{p^2}{m_i^2}\gg1$)
\begin{equation}N_i\approx1-\frac{\frac{p^2}{2m_i^2}\sin^2\phi}{1+\frac{p^2}{m_i^2}\sin^2\phi}
\approx\frac{m_i^2}{p\sin^2\phi}.\end{equation}
Thus, for typical combinations of the coefficients $|c_i^\pm|^2$ of eq.(\ref{P}) in the linear approximation over $\cos \phi \ll 1$ we get

  \begin{equation}|c_i^+|^2|c_i^-|^2=\frac{1-N_i^2\cos^2\phi}{4}\approx\frac{1}{4},\end{equation}
  \begin{equation}|c_2^s|^2|c_1^s|^2\approx\frac{1}{4}\left(1-s(N_1+N_2)\cos\phi\right)
       \approx
       \frac{1}{4}\left(1-s\frac{m_1^2+m_2^2}{p^2}\cos\phi\right), \end{equation}
\begin{equation}|c_2^+|^2|c_1^-|^2\approx\frac{1}{4}\left(1-(N_2-N_1)\cos\phi\right)\approx
  \frac{1}{4}\left(1-\frac{m_2^2-m_1^2}{p^2}\cos\phi\right),\end{equation}
\begin{equation}|c_2^-|^2|c_1^+|^2\approx\frac{1}{4}\left(1-(N_1-N_2)\cos\phi\right)\approx
  \frac{1}{4}\left(1-\frac{m_1^2-m_2^2}{p^2}\cos\phi\right).\end{equation}
Using these expressions finally from \eqref{P} we get
\begin{multline}\label{fin}
  P_{\nu_{eL}\rightarrow\nu_\mu}(t)=\frac{1}{4}\sin^22\theta\left\{-\sin^2\frac{E_2^+-E_2^-}{2}t+
  \sin^2\frac{E_2^+-E_1^+}{2}t+\sin^2\frac{E_2^+-E_1^-}{2}t+
  \right.\\\left.+\sin^2\frac{E_2^--E_1^+}{2}t+\sin^2\frac{E_2^--E_1^-}{2}t
  -\sin^2\frac{E_1^+-E_1^-}{2}t\right\}+\\+\frac{m_1^2+m_2^2}{p^2}\sin^22\theta\cos\phi\left\{
  \sin^2\frac{E_2^--E_1^-}{2}t-\sin^2\frac{E_2^+-E_1^+}{2}t\right\}+\\+\frac{m_2^2-m_1^2}{p^2}\sin^22\theta\cos\phi\left\{\sin^2\frac{E_2^--E_1^+}{2}t-\sin^2\frac{E_2^+-E_1^-}{2}t\right\}.
\end{multline}
Note that two last terms here are suppressed by the presence of $\cos \phi \ll 1$.

If we also account for a rather general condition $2\mu B_\bot \ll p$  then the following approximation can be used for the neutrino energies
\begin{equation}E_i^\pm\approx\sqrt{m_i^2+p^2\pm2\mu_{i} pB_\bot\left(1+\frac{m_i^2}{2p^2}\right)}\approx
p\sqrt{1+\frac{m_i^2}{p^2}\pm\frac{2\mu_{i} B_\bot}{p}}\approx p+\frac{m_i^2}{2p}\pm\mu_{i} B_\bot.\end{equation}
Finally,  for the neutrino oscillations probability in the flavour basis we get
\begin{multline}\label{appP}
  P_{\nu_{eL}\rightarrow\nu_\mu}(t)\approx\sin^22\theta\sin^2\frac{\Delta m^2}{4p}t+\frac{1}{2}\left(\sin^2\frac{\mu_2-\mu_1}{2}
  B_\bot t+\sin^2\frac{\mu_2+\mu_1}{2}B_\bot t\right)
  \sin^22\theta \cos\frac{\Delta m^2}{2p}t \\-\frac{1}{4}\sin^22\theta\left(\sin^2\mu_1B_\bot t+
  \sin^2\mu_2B_\bot t\right),
\end{multline}where $\Delta m^2\equiv m_2^2-m_1^2.$

It should be emphasized that, as it follows from the above derivations, the obtained probability $P_{\nu_{eL}\rightarrow\nu_\mu}(t)$ accounts for the transitions from the initial
left-handed electron neutrino to the final muon neutrino that can be in both left- and right-handed
states.  Note that in the mass basis the transition $\nu_{1 L}\rightarrow\nu_{2 R}$ is not possible when $\mu_{12}=0$ (see eq.(\ref{2_evol_eq})).

Consider the difference $\Delta P(t)=P_{\nu_{eL}\rightarrow\nu_\mu}-P_{\nu_{eL}\rightarrow\nu_{\mu L}}$ of the probability $P_{\nu_{eL}\rightarrow\nu_\mu}(t)$ given by (\ref{appP})
and the usual result
 \begin{equation}\label{compare}
P_{\nu_{eL}\rightarrow\nu_{\mu L}}(t)=\sin^22\theta\sin^2\frac{\Delta m^2}{4E_\nu}t
\end{equation}
for the neutrino flavour transition $\nu_{eL}\rightarrow\nu_{\mu L}$.
Obviously, this difference is just a probability for the transition $P_{\nu_{eL}\rightarrow\nu_{\mu R}}$, and
we get
\begin{multline}\label{appP}
  P_{\nu_{eL}\rightarrow\nu_{\mu R}}=\frac{1}{2}\left(\sin^2\frac{\Delta\mu_-}{2}
  B_\bot t+\sin^2\frac{\Delta\mu_+}{2}B_\bot t\right)
  \sin^22\theta \cos\frac{\Delta m^2}{2p}t \\-\frac{1}{4}\sin^22\theta\left(\sin^2\mu_1B_\bot t+
  \sin^2\mu_2B_\bot t\right),
\end{multline}
where the notations are used $\Delta\mu_\pm=\mu_2\pm\mu_1.$

Here we recall that the result (\ref{appP}) is obtained within an approach that implies the description of the neutrino spin states by using the spin operator (\ref{spin}) that commutes with the neutrino dynamical Hamiltonian. Contrary to what is usually obtained (see, for instance, in \cite{Giunti:2014ixa})  that the
transitions $\nu_{eL}\rightarrow\nu_{\mu R}$ in $B_{\perp}$ are not possible in the
case of zero transition magnetic moments, from (\ref{appP}) we show that the discussed transitions
can proceed even in the case when the transition magnetic moments are zero.
This predicted new phenomenon can have important phenomenological consequences in implications of neutrino transitions in magnetized astrophysical media.

\section{Acknowledgements}
One of the authors (A.S.) is thankful
to George Zoupanos and Margarida Rebelo for the kind invitation to stay at the Corfu Summer Institute for the Summer School and Workshop on the Standard Model and Beyond. This study has been partially supported by the Russian Foundation for Basic Research (grants No. 14-22-03043-ofi and No. 15-52-53112).

\selectlanguage{english}

\end{document}